\documentclass{emulateapj}
\usepackage{graphicx}

\usepackage{natbib}
\def\etal  {et~al.}
\shortauthors{Fixsen} \shorttitle{CMB Temperature}
\tighten

\begin{document}
\def\COBE{{\sl COBE\/}}
\def\wisk#1{\ifmmode{#1}\else{$#1$}\fi}
\def\um     {\wisk{{\rm \mu m}}}
\def\etal   {et~al.~}
\def\deg    {\wisk{^\circ}}
\def\tf    {2.72548}
\def\icm    {\wisk{\rm cm^{-1}}}
\hfuzz=10pt \overfullrule=0pt
\pretolerance=1000  

\title{The Temperature of the Cosmic Microwave Background}

\author{D.~J.~Fixsen\altaffilmark{1}}

\altaffiltext{1}{University of Maryland, Goddard Space Flight
Center}

\begin{abstract}
The FIRAS data are independently recalibrated using the WMAP data to
obtain a CMB temperature of $2.7260\pm 0.0013$. Measurements of the
temperature of the cosmic microwave background are reviewed. The
determination from the measurements from the literature is cosmic
microwave background temperature of \tf$\pm 0.00057$~K.
\end{abstract}


\section{Introduction}
The Wilkinson Microwave Anisotropy Probe (WMAP) data presents an
opportunity to recalibrate the Far InfraRed Absolute
Spectrophotometer (FIRAS) experiment and produce an independent
check of the other measurements of the cosmic microwave background
(CMB) temperature. In sections 2~\&~3 the WMAP data will be
presented and combined with the FIRAS data to make an independent
estimate of the CMB temperature. In sections 4~\&~5 this new
estimate will be combined with others from the literature to
generate an improved estimate for the CMB temperature.

The WMAP data only measures the difference in intensity between
different points on the sky. However, the precision is sufficient
such that the velocity of the WMAP spacecraft can be used to
calibrate the velocity to various points on the surface of last
scattering of the CMB. This velocity in turn is used to form a
differential spectrum of the CMB. The differential spectrum is then
fit with a single parameter which is the CMB temperature.

\section{WMAP Velocity Map}
The standard WMAP sky maps (Hinshaw \etal 2009) are corrected to the
baricenter of the solar system using the JPL ephemeris (Standish \&
Fienga 2002). The calibration assumes a CMB temperature of 2.725~K
(derived from the FIRAS measurement). However, the various changes
as the WMAP makes its way around the sun (now in its ninth
repetition) can be used to calibrate the WMAP data in terms of
velocity. The velocity of the WMAP spacecraft, with respect to the
sun, is known to $<1$~cm/s, which is a negligible uncertainty
compared to other uncertainties considered here. This velocity is
used to calibrate a map of velocity relative to the surface of last
scattering of the CMB. Most of this velocity is the dipole,
presumably the motion of the solar system with respect to the frame
of the CMB, however it includes temperature variations due to the
Sachs-Wolfe effect which has the same spectrum. This process is
repeated for each WMAP differential assembly and each year yielding
50 independent maps for the first 5 years of WMAP operation.

The WMAP velocity maps are the temperature maps, available at
http://lambda.gsfc.nasa.gov/ product/map/current/m\_products.cfm,
with the dipole added back in and divided by the CMB temperature. In
generating the WMAP temperature maps a dipole is fit and removed
from the raw data. The residual variations are due to various
sensitivities of the WMAP instrument, the changing velocity as the
spacecraft makes its annual trek around the solar system and the
small variations of the CMB as a function of position. The WMAP team
has done an excellent job of removing the instrumental effects. By
correcting the velocity to the barycenter of the solar system, the
effects of the spacecraft velocity are also removed. But this
information also allows the calibration of the WMAP data. But the
information is in velocity rather than temperature. In order to get
the temperature for the published maps an absolute temperature must
be assumed. The WMAP team used 2.725~K from the previous FIRAS
measurements to translate the velocity measurements into temperature
units. To get the velocity maps, the data is divided by 2.725~K to
restore the velocity calibration in units of $v/c$. The dipole
$(v/c=0.0012338,l=263.87^\circ,b=48.24^\circ)$ is added back in to
restore the full velocity. The fitting includes an absolute term
which is treated as a nuisance parameter so no absolute adjustment
to the WMAP data is required or made.

No corrections to the WMAP data are made for Galactic foregrounds.
The maps are convolved with the FIRAS beam (approximately a 7 deg
tophat) to produce maps with the FIRAS resolution. These maps are
produced with the pixelization of the native FIRAS data.

\begin{figure*}[t]
\includegraphics[trim=0 0 0 0,width=7in]{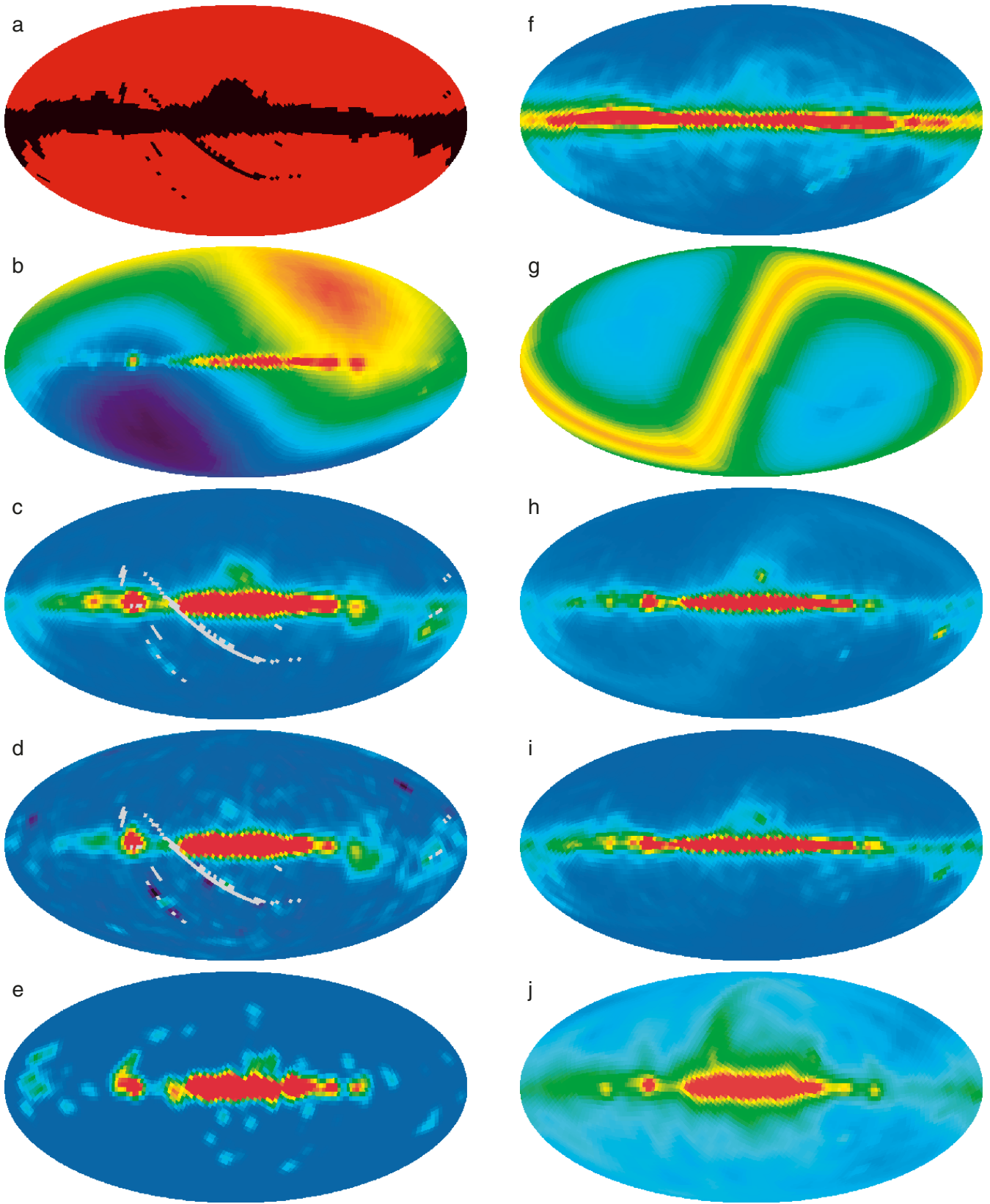}
\caption{The templates used to fit the FIRAS data. The templates are
in Galactic coordinates with the center of the Galaxy in the center
of each Mollwide projection. The Galactic equator is a horizontal
line across the middle of each figure. From top to bottom the
templates are a)Data mask b)WMAP velocity c)C{\sc ii} d)N{\sc ii}
e)Al26 f)H{\sc i} g)Zodical model h)DIRBE band 8 i)DIRBE band 10
j)Haslam 408 MHz.} \label{Templates}
\end{figure*}

For each velocity map, the FIRAS data (also available on the lambda
website) are fit to a set of ten templates, shown in figure
\ref{Templates}. The first template is unity everywhere. This
template is included to model the monopole. The second template is
one of the velocity maps from the WMAP data. The remaining eight
maps are included to fit various foregrounds. These templates
include the FIRAS C{\sc ii} and N{\sc ii} maps (Fixsen \etal\ 1999).
The Band 8 and 10 DIRBE data also from the COBE mission and a model
of the zodiacal emission from the DIRBE team (Kelsall \etal 1998).
Also included are an H{\sc i} map (Dickey \& Lockman 1990, Hartmann
\& Burton 1997), an Aluminum-26 emission line map (Diehl \etal\
1995) and the Haslam 408 MHz map (Haslam \etal\ 1981). The maps not
derived from FIRAS data are convolved with the FIRAS beam. Templates
3-10 are included to attempt to fit all local features even though
many subsets of them would do almost as well. Since there are 6063
pixels the 8 templates make an insignificant reduction in the total
number of degrees of freedom.

The FIRAS data is arbitrarily cut off at 1~THz since the CMB
spectrum is essentially zero beyond that frequency and the
uncertainty of the FIRAS data at high frequencies is not important
for this analysis. The fits use the weight map of the FIRAS low
frequency data (which is nearly identical in form to the high
frequency weight map). Each frequency is fit separately and so each
template generates a spectrum for each fit. All of the spectra
except the spectrum associated with the velocity are treated as
nuisance parameters. These increase the uncertainty but are
otherwise unused.

\begin{figure}[t]
\includegraphics[angle=90,trim=0 0 10 50,width=3.5in]{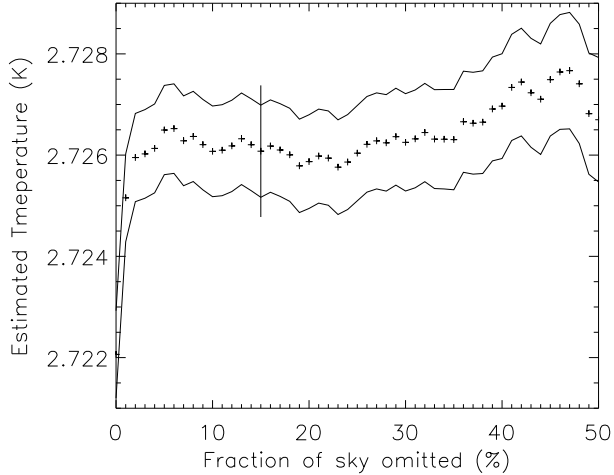}
\caption{The crosses are the CMB temperature estimation for a given
fraction of the brightest part of the sky excluded. The lines are
the nominal +1~$\sigma$ and -1~$\sigma$ limits. The error bar is the
adopted value and uncertainty inflated for the excess $\chi^2$.}
\label{cut}
\end{figure}

Even with these templates the Galactic plane is too large a
perturbation to be insignificant. So the pixels with the largest
DIRBE band 10 signal are excluded from the fit. The process was
tested with 0\% to 90\% of the data excluded. Figure \ref{cut} shows
the average estimated temperature for all 50 WMAP channels and years
as a function of the fraction of excluded data. For less than 5\% of
the data excluded, significant variations are seen. The variations
are concentrated in the the 22~GHz (K band) templates. If more than
50\% of the data are excluded the uncertainties are significantly
larger, not only due to the data loss but also due to the loss of
the hot and cold ends of the CMB dipole. The results that follow are
for 15\% of the data excluded. These fits, done independently at
each frequency, result in 50 spectra.

\section{CMB Velocity Temperature}

\begin{figure*}[t]
\includegraphics[angle=90,trim=0 20 10 50,width=7in]{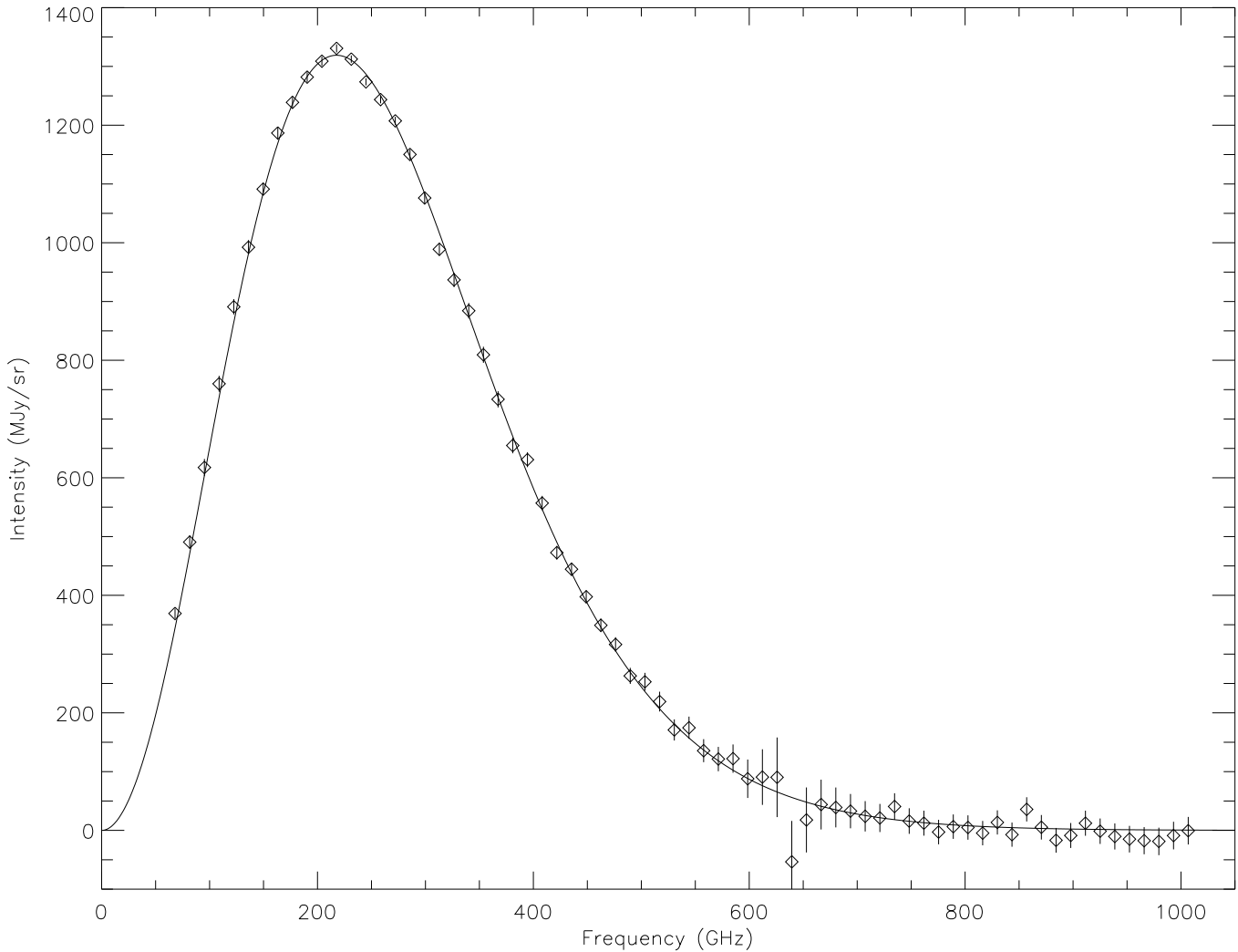}
\caption{The mean spectrum associated with the velocity of the solar
system with respect to the CMB. The line is the {\it a priori}
prediction based on the WMAP velocity and the previous FIRAS
calibration. The uncertainties are the noise from the FIRAS
measurements. The error bars are slightly misleading, because they
do not show the correlations, but the correlated errors are properly
treated in the fit.} \label{FIRAS}
\end{figure*}

The average of the 50 spectra is shown in fig. \ref{FIRAS}. Since
this spectrum is associated with the velocity of the solar system
with respect to the CMB (and some Sachs-Wolfe effect), the spectrum
appears not as an absolute spectrum but as a differential spectrum
generated by the doppler shift.

According to the special theory of relativity the spectrum observed
in a reference frame moving with respect to the source of a
black-body spectrum, $B(T,\nu)$, at temperature $T$, is shifted such
that
\begin{equation}
S'(\nu)=B(T\sqrt{{1+v}\over{1-v}},\nu)
\end{equation}
where $\nu$ is the frequency and $v$ is the velocity {\it towards}
the source divided by the speed of light.

Thus ignoring negligible second order terms and higher we have:
\begin{equation}
S(\nu,p)=B(T,\nu)+v(p)T{\partial B(T,\nu)\over \partial T} |_{T=T_0}
\end{equation}
where the spectrum is now a function of position, $P$, on the sky.
Note that the first term is absorbed in the first template. The
remaining term, $vT\partial B/\partial T$ is the term related to the
2nd template. The velocity, $v$, is already included in the
template, thus the term to fit is $T\partial B/\partial T$, which
must match the spectrum. Although strictly speaking this is
nonlinear, a spectrum $T\partial B/\partial T$ for a $T$ near the
CMB temperature (say 2.726~K) can be subtracted from the spectrum
and the residual can then be fit to $\delta T(\partial B/\partial
T+T\partial^2 B/\partial T^2)$ for the small $\delta T$ correction.
Since both the process of averaging the velocity maps and the
spectral fitting are linear processes, they can be done in either
order. So the average spectrum is fit with a result of $T=2.7260$.

\subsection{FIRAS Uncertainty}
The uncertainty of the temperature is dominated by the noise in the
FIRAS measurements. Propagating the uncertainties shown in figure
\ref{FIRAS} results in an uncertainty estimate of 0.74~mK, but this
does not include the correlations amongst the different frequencies.
Including the correlations and the PEP error term, which is
important in this context (Fixsen \etal\ 1994), results in an
uncertainty estimate of 1.09~mK. The $\chi^2$ is 98.7 for 69 DOF.
Since the $\chi^2$ is higher than expected the uncertainty is
inflated to produce a $\chi^2$ per DOF of unity with a resulting
uncertainty of 1.3~mK in the CMB temperature due to the
uncertainties of the FIRAS measurements.

In deriving the FIRAS dipole (Fixsen \etal\ 1996) amplitude the only
term is the $\delta T \partial B/\partial T$. Here the second term
$\delta T T\partial^2 B/\partial T^2$ is both larger and at a mean
higher frequency. This allows a more precise determination of the
temperature. Also the velocity map allows more control of the
systematic effects from the cut. Some of the variation due to the
cuts was due to higher order ($l>1$) variation in the CMB. Here the
variation in the CMB is included in the velocity maps, so neither of
these terms add to the uncertainty.

\subsection{WMAP Uncertainty}
The uncertainty from the WMAP measurements also needs to be
included. The RMS variation in the gains ($\Delta v/v$) of the 50
independent WMAP maps is 0.00096. Nominally, the uncertainty of the
mean of such a set would be 0.00014. Figure 7 of Hinshaw \etal\
(2009) indicates a .001 uncertainty for each of 40 individually
calibrated data channels which results in a similar uncertainty
estimate for the mean.

But at the level of 0.00014, the WMAP uncertainties are
insignificant. Even if the uncertainty were 0.0005 the WMAP data
would not the be a significant source of uncertainty. If the mean
error were 0.0016, the WMAP uncertainty would equal the uncertainty
from the FIRAS uncertainty and the final uncertainty would be
increased by a factor of $\sqrt{2}$. Each WMAP channel and year is
processed independently. Unless there is a serious unexplained error
in the WMAP data that correlates both channels and years the
uncertainty of the WMAP data is insignificant relative to the
uncertainty in the FIRAS data.

The fitted temperatures for each WMAP channel and year is shown in
table 1. The average is 2.7260~K with a standard deviation of
0.6~mK. The implied uncertainty of the mean is 0.09~mK. The standard
deviation is smaller than expected because a significant part of the
variation of the WMAP velocity maps is concentrated in the Galactic
plane. The 0.09~mK does not include the common uncertainty of the
WMAP velocity maps or the FIRAS errors.

\begin{table}[tbph]
\begin{center}
\begin{tabular}{rccccc}
WMAP & Year 1 & Year 2 & Year 3 & Year 4 & Year 5  \\\hline
 K  & 2.72511 & 2.72517 & 2.72522 & 2.72512 & 2.72519 \\
 Ka & 2.72532 & 2.72530 & 2.72526 & 2.72529 & 2.72544 \\
 Q1 & 2.72555 & 2.72541 & 2.72540 & 2.72546 & 2.72557 \\
 Q2 & 2.72553 & 2.72548 & 2.72538 & 2.72543 & 2.72544 \\
 V1 & 2.72625 & 2.72627 & 2.72621 & 2.72620 & 2.72639 \\
 V2 & 2.72653 & 2.72622 & 2.72630 & 2.72626 & 2.72662 \\
 W1 & 2.72624 & 2.72616 & 2.72644 & 2.72635 & 2.72634 \\
 W2 & 2.72681 & 2.72682 & 2.72656 & 2.72665 & 2.72650 \\
 W3 & 2.72675 & 2.72678 & 2.72649 & 2.72686 & 2.72714 \\
 W4 & 2.72712 & 2.72685 & 2.72617 & 2.72580 & 2.72658 \\
\end{tabular}
\caption{\label{temps} Estimated CMB temperatures for the various
years and WMAP channels. All of these fits are with the brightest
15\% of the sky excluded.}
\end{center}
\end{table}

Combining the uncertainties of the FIRAS and WMAP in quadrature
results in an uncertainty of 1.3~mK. This measurement is unaffected
by absolute systematic errors of either FIRAS or WMAP as it uses
only differential measurements of both experiments. Further it is
insensitive to long term offset drifts in either instrument as the
measurement is dominated by the measurement at the precession period
of the WMAP data (about 60 minutes) and the orbital period of the
COBE spacecraft (about 100 minutes).

In particular it does not depend on the absolute calibration of the
FIRAS thermometers although it does depend on the gain calibration
of the FIRAS external calibrator thermometers. It also is relatively
insensitive to the FIRAS frequency calibration as the measurement
uncertainty is dominated by the amplitude of the spectrum.

The diamonds in fig \ref{FIRAS} are the spectra derived from the
FIRAS data. The uncertainties shown are also derived from the FIRAS
data. Note the line in figure \ref{FIRAS} is not a fit but the
predicted line from the previous T=2.725 FIRAS calibration. The best
fit is 2.7260$\pm$0.0013~K. The feature at $\nu=630$~GHz, including
the larger uncertainty, is due to the dichroic filter separating the
low and high frequencies in the FIRAS instrument.

\section{CMB Temperature}
There were many publications of measurements of the CMB temperature
from the late '60s and '70s.  But the uncertainties are large and
the systematics were not well understood. Here an arbitrary cutoff
of 50~mK uncertainty was used to select 15 sources for the CMB
temperature from recent publications. These results are shown in
Table 2.

\begin{table}[tbph]
\begin{center}
\begin{tabular}{lccl}
CMB  & Temp & Uncertainty &
\\ Source & (K) & (mK) & Reference \\\hline
CN &     2.700 & 40 & Meyer \& Jura (1985)\\
CN &     2.740 & 50 & Crane \etal\ (1986)\\
Balloon& 2.783 & 25 & Johnson \& Wilkinson (1987)\\
CN &     2.750 & 40 & Kaiser \& Wright (1990\\
Rocket & 2.736 & 17 & Gush \etal\ (1990)\\
S Pole & 2.640 & 39 & Levin \etal\ (1992)\\
Balloon& 2.712 & 20 & Schuster \etal\ (1993)\\
CN &     2.796 & 39 & Crane \etal\ (1994)\\
CN &     2.729 & 31 & Roth \etal\ (1995)\\
Balloon& 2.730 & 14 & Staggs \etal\ (1996)\\
ARCADE1& 2.694 & 32 & Fixsen \etal\ (2004)\\
ARCADE1& 2.721 & 10 & Fixsen \etal\ (2004)\\
ARCADE2& 2.731 &  5 & Fixsen \etal\ (2009)\\
FIRAS & 2.7249 & 1.0 & Mather \etal\ (1999)\\
FIRAS & 2.7255 & 0.85 & Fixsen \etal\ (1996)\\
FIRAS & 2.7260 & 1.3 & This Work\\
\hline Mean & \tf & .57 &\\\hline
\end{tabular}
\caption{\label{temps} Measurements and uncertainties of the CMB
temperature.}
\end{center}
\end{table}

The weight of the measurements is dominated by the FIRAS
measurements, but it is still instructive to look at the other
measurements.  The measurements using CN are entirely different from
any of the other measurements. But the combined CN measurements are
2.742$\pm$0.017~K which is only one sigma high. The rocket
measurement from Gush, Halpern and Wishnow (1990) is like the FIRAS
measurement in that it uses a fourier transform spectrometer, but it
has an independent calibrator and independent thermometers. The
other measurements depend on external calibrators as the absolute
reference, but these are comparatively narrow bands on the low
frequency side of peak of the CMB radiation. The $\chi^2$ for the 16
measurements is 17.9 for 15 DOF.  A $\chi^2$ this large should be
expected about 26\% of the time. Most of the excess $\chi^2$ comes
from three measurements: the Johnson \& Wilkinson balloon
measurement, the South Pole measurement, and the fourth CN
measurement. With this number of measurements, one or two 2$\sigma$
results should be expected but here there are three.

Without the FIRAS measurements the weighted average is
2.729$\pm$0.0038 which is $1.0~\sigma$ from the final answer. Most
of the weight (97\%)of the final temperature estimate is from the
FIRAS measurements. Each FIRAS measurement will be reviewed in turn.

The original concept of the FIRAS instrument was that the sky would
be observed and the internal reference would be adjusted to minimize
the signal. Then the external calibrator would be inserted and
adjusted to match the signal from the sky. This method depends on
knowing the calibration of the external calibrator germanium
resistance thermometers. A 5 mK error in the original temperature
determination of the external calibrator led directly to a 5 mK
error in the temperature determination of the CMB. However, the
calibration process corrects other effects of the error to first
order (Fixsen \etal 1994). There were slight modifications to this
plan (eg. the internal reference was offset 10 mK for about half of
the data), but basically the result is 2.7249~K (This has been
rounded off to 2.725 in the literature), including the 5~mK
correction for thermometer self-heating in the high current mode.
The low current mode is noisy but there are $\sim 10^5$ measurements
to compare the low current readings to the high current readings all
at $\sim 2.7$~K. The final uncertainty depends on the calibration of
the thermometers at NIST and the readout electronics.
The uncertainty is estimated as 1~mK.

The second calibration of the FIRAS data is based on the color. The
temperature can be determined from the color of the radiation if the
frequency scale can be accurately determined. The frequency scale is
derived from FIRAS observations of the interstellar CO and [C~I]
lines at 1300, 867, 650, and 609 \um\ (Fixsen \etal 1996). These
were chosen because they are bright enough to determine the
frequency and they are in the same part of the spectrum as the CMB.
The temperature scale was determined independently from 7 different
combinations of the four detectors and three scan modes. These
determinations agreed within their uncertainties and the weighted
uncertainty is 0.2 mK. There is a common uncertainty of 0.82 mK due
to the uncertainty of the frequency scale, which dominates the total
uncertainty. The result is 2.7255~K~$\pm$0.85~mK. The thermometer
errors are only weakly coupled to the color temperatures. Indeed,
when the first color temperature was published, it disagreed with
the thermometer by 4.5~$\sigma$. It was only with the discovery of
the high current self-heating offset that the two measurements came
into alignment. The uncertainty of this method is driven by the
uncertainty in the measurement of the frequency of the CO and [C~I]
lines.


Now there is a third independent method of precisely calibrating the
FIRAS instrument. The velocity method was presented before with the
COBE DMR data (Fixsen \etal\ 1996). However, the DMR data did not
have sufficient velocity precision to fully exploit this method; the
WMAP data do. Because the differential measurements from the FIRAS
instrument are taken only 50~minutes (half of an orbit) apart with
the instrument in substantially the same state this method has the
least potential for systematic errors. These spectra have far more
supporting data than the calibration data. Each of the WMAP
frequencies can be used to construct velocity map which in turn can
be used to construct a spectrum. These spectra can then be fit to a
dB/dT spectrum with the temperature as the single adjustable
parameter.

\section{SUMMARY AND CONCLUSIONS}

The calibration methods for the Far Infrared Absolute
Spectrophotometer (FIRAS) have been described and the accuracy
estimated. All of the recent precision estimates of the CMB
temperature agree within 2.5 times their uncertainties. These
estimates were made with a variety of methods from different
platforms and different frequencies. Combining all of the estimates
results in a very modestly elevated $\chi^2$ and an improved
absolute temperature estimation of $\tf\pm 0.00057$~K.

\acknowledgements

I thank the WMAP team for providing the smoothed sky maps in
velocity units. A special thanks to J Weiland and G Hinshaw.

\end{document}